\documentclass[10pt]{iopart}
\usepackage [dvips]{graphicx,psfrag}

\newcommand{\beq}{\begin{equation}}
\newcommand{\eeq}{\end{equation}}
\newcommand{\tc}{T_c}
\newcommand{\jpsi}{$J/ \psi$}
\newcommand{\ec}{$\eta_c$}
\newcommand{\chc}{$\chi_{c_i}$}
\newcommand{\nt}{$N_\tau$}  
\newcommand{\sg}{\sigma(\omega,T)}
\newcommand{\sgh}{\sigma_H(\omega,T)}
\newcommand{\rhw}{$\rho(\omega,T)$}
\newcommand{\xxx}{{\em Preprint}}
\newcommand{\om}{\omega}
\newcommand{\grecon}{G_{\rm recon, T^*}(\tau, T)}
\newcommand{\Gt}{G(\tau, T)}
\newcommand{\ax}{$\chi_{c_1}$}
\newcommand{\as}{$\chi_{c_0}$}
\begin{document}

\title{Charmonium systems after the deconfinement transition}

\vspace*{-3cm}
\mbox{} \hfill BI-TP 2004/07, DESY 04-049\\
\vspace*{2cm}

\author{Saumen Datta,\dag\ Frithjof Karsch,\dag\ Peter
Petreczky\ddag\ \footnote[4]{Goldhaber and RIKEN fellow} 
and Ines Wetzorke\S }

\address{\dag\ Fakult\"at f\"ur Physik, Universit\"at Bielefeld, 
D-33615 Bielefeld, Germany}

\address{\ddag\ Physics Department, 
Brookhaven National Laboratory, Upton, NY 11973, USA}

\address{\S\ NIC/DESY Zeuthen, Platanenallee 6, D-15738 Zeuthen, Germany.}

\begin{abstract}
The behavior of charmonia after the deconfinement transition
is investigated on quenched
lattices. Analysis of temporal correlators on fine lattices
at temperatures upto 3 $\tc$ show that the \jpsi\ and
\ec\ survive the deconfinement transition with little significant
changes, and survive as bound states at least upto 2.25 $\tc$. The
spatially excited \chc\ states suffer serious system
modifications, maybe dissolution, already a little above $\tc$.
\end{abstract}




The behavior of charmonia in heavy ion collisions has received
considerable attention ever since Matsui \& Satz 
suggested absence of \jpsi\ bound states as a signal of deconfinement,
based on nonrelativistic potential model arguments  
\cite{matsui}. A different, more dynamical line of argument 
supported this suggestion, showing that the hard gluons available in
the deconfined plasma will readily break the \jpsi\
(treating \jpsi\ as a Coulombic state) \cite{kharzeev}.
Direct studies on quenched lattices, however, show a quite 
different picture: the 1S states like \jpsi\ and \ec\
survive the deconfinement transition with little
significant change, and bound states survive in
gluonic plasma at least upto temperatures of 1.5 $\tc$ 
\cite{lat02,matsufuru,asakawa1}. Here we report an update of our
results in reference \cite{lat02}.
Further details, including discussion of systematics, 
can be found in \cite{paper}. 

Lattice studies of finite temperature QCD employ the
Matsubara formalism, in which field theory at thermal 
equilibrium is studied by trading the time axis for the 
inverse temperature. The detailed properties of mesonic states
can be obtained from the Matsubara correlators
\beq
G_H (\tau, T) = \langle J_H (\tau) J_H^\dag
(0) \rangle_T.
\label{eq.corr} \eeq
Here $J_H$ is the suitable hadronic operator, $\tau$
denotes the Euclidean time and we consider
zero momentum operators only. 
An integral equation then connects $G_H(\tau, T)$ to
the spectral function, $\sgh$, for the operator:
\beq
G_H(\tau, T) = \int_0^{\infty} d \omega
\sgh \frac{\cosh(\omega(\tau-1/2 T))}{\sinh(\omega/2 T)}.
\label{eq.spect}
\eeq

The high temperatures that are of interest to us correspond
to a small temporal extent. In order to have a reasonable number of
data points for the correlator, we use very fine
lattices. We present results here for two sets of
lattices, with lattice spacings $\sim$ 0.02 and 0.04 fm,
respectively, using the finer set to reach higher temperatures
and the coarser set for temperatures closer to $\tc$.
For each set, we use a quark mass close to the charm ($m_{J
/ \psi}$ $\sim$ 3.1 GeV for the coarser set and $\sim$ 3.7
GeV for the finer set), and change the time extent \nt\ to vary the
temperature. Details of the lattices studied and the
corresponding temperatures are tabulated below.

\begin{center}\begin{tabular}{ccccc|ccccc}
\hline
a[fm] & $N_{\sigma}^3 \times N_{\tau}$ & T/$\tc$ & \#
configs. & &
a[fm] & $N_{\sigma}^3 \times N_{\tau}$ & T/$\tc$ & \#
configs. \\
 & $48^3 \times 24$ & 0.75 & 100 & &
 & $40^3 \times 40$ & 0.9 & 85 \\
0.04 & $48^3 \times 16$ & 1.12 & 50 & &
0.02 & $64^3 \times 24$ & 1.5 & 80 \\
& $48^3 \times 12$ & 1.5 & 60 & &
& $48^3 \times 16$ & 2.25 & 100 \\
& & & & & & $48^3 \times 12$ & 3.0 & 90 \\
\hline
\end{tabular} \end{center}

The ill-defined problem of inverting \Eref{eq.spect} to
extract $\sg$ can be handled with the ``maximum entropy
method'' (MEM) \cite{asakawa2}, which provides a prior guess for the
solution through a Shannon-Jaynes ``entropy'' 
term(see \cite{jarrel} for a review). The application of this 
method at finite temperature has led to important
qualitative information about the
dilepton yield from an equilibriated plasma \cite{ines}. 
However, the small extent of
the Euclidean time direction at finite temperature 
makes the problem even more difficult \cite{matsufuru}, and
the role of the prior information becomes important (see
\cite{paper} for details). 

\begin{center} \begin{figure} [htb]
\centerline{\includegraphics[width=6.2cm]{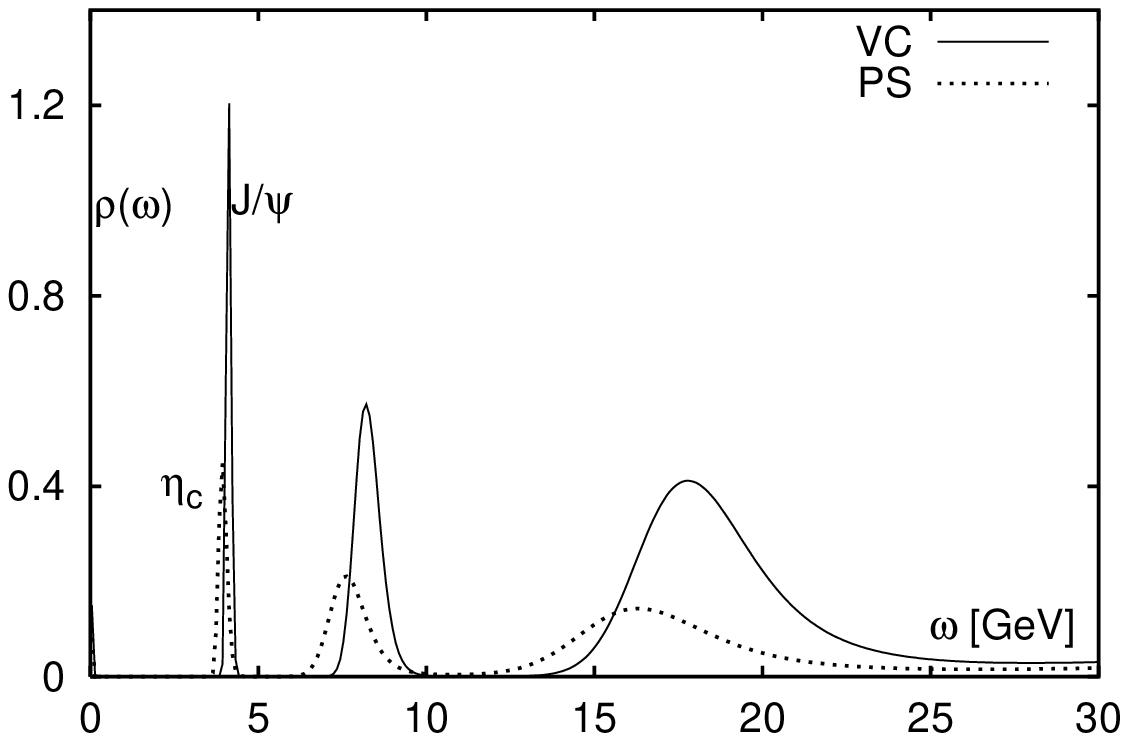}
\includegraphics[width=6.2cm]{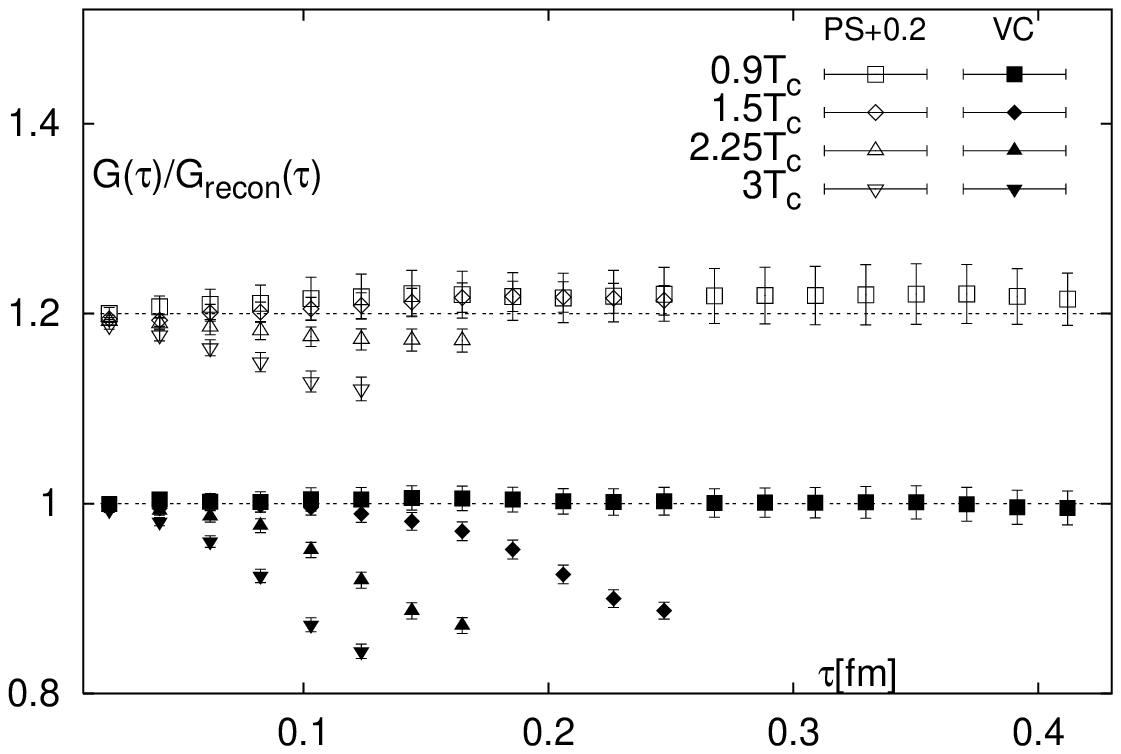}}
\caption{a) The spectral function extracted from the
temporal correlators at 0.9 $\tc$ 
in pseudoscalar and vector channels. The 
first peaks correspond to \ec\ and \jpsi\,
respectively. 
b) $\Gt /\grecon$, where $\grecon$ is constructed
from the spectral functions in a) (see text).}
\label{fig.psvc0.9}
\end{figure} \end{center}

The pseudoscalar and vector states are explored by using the 
point-point operators $J_H$ = $\bar{c} \gamma_5 c$ and
$\bar{c} \gamma_\mu c$, respectively. The lightest states
in these channels correspond to \ec\ and \jpsi\
respectively. The extraction of the spectral
functions below $\tc$, using \Eref{eq.spect} and applying 
the MEM procedure, is
stable and reproduces the properties of the ground state
reasonably well. The spectral functions for these states
at 0.9 $\tc$ (for the lattice with lattice spacing 0.02 fm;
see table above) are shown in  \Fref{fig.psvc0.9}
a). Here we plot \rhw\ = $\sg / \om^2$.
The ground state peaks and their strengths reproduce
reasonably well the mass and amplitude obtained from a fit.
The peaks at higher $\om$ scale
roughly with the lattice spacing and are probably dominated by
lattice artefacts (see the discussion in reference
\cite{paper}). 

\vspace*{-0.4cm}
\vspace*{-0.4cm}

As a first step towards looking for the effect of the
deconfinement transition on the \jpsi\ , one can ask the
question whether the spectral functions in
\Fref{fig.psvc0.9} a) explain the correlators above $\tc$. 
We construct the model
correlators above $\tc$ according to \Eref{eq.spect},
\beq
\grecon = \int_0^{\infty} d \omega
\sigma(\om, T^*) \frac{\cosh(\omega(\tau-1/2
T))}{\sinh(\omega/2 T)}
\label{eq.recon}
\eeq
where $T^*$ is a temperature below $\tc$, in this case 0.9
$\tc$. A comparison of $\grecon$ with the directly measured
correlators $\Gt$ give an indication of
temperature modification of the mesonic properties above
the transition. 

This simple comparison turns out to reveal a lot about the
properties of \jpsi\ and \ec\ above the
transition. \Fref{fig.psvc0.9} b) shows the result of
such a comparison. For the \ec\
channels, the spectral function at 0.9 $\tc$ is seen to
completely explain the measured correlators at 1.5 $\tc$,
indicating that there is no significant change in this
channel upto this temperature. At 2.25 $\tc$, $\Gt$ shows
only small deviations from $\grecon$, while at 3 $\tc$
significant changes are seen. For the vector channel,
at 1.5 $\tc$ $\Gt$ is described by $\grecon$ at small distances, while
small deviations are seen at larger distances. The deviations appear 
at shorter distances at 2.25 $\tc$, while pretty large deviations are seen
at all distances at 3 $\tc$.

\begin{center} \begin{figure} [htb]
\centerline{\includegraphics[width=6.2cm]{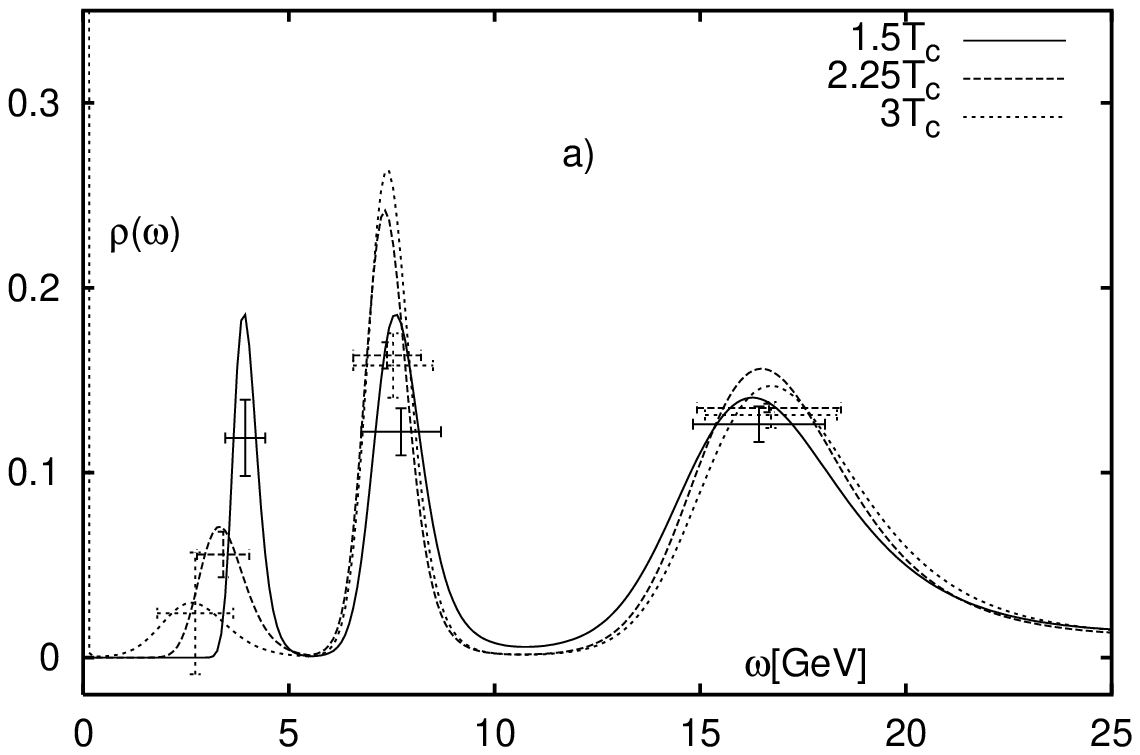}
\includegraphics[width=6.2cm]{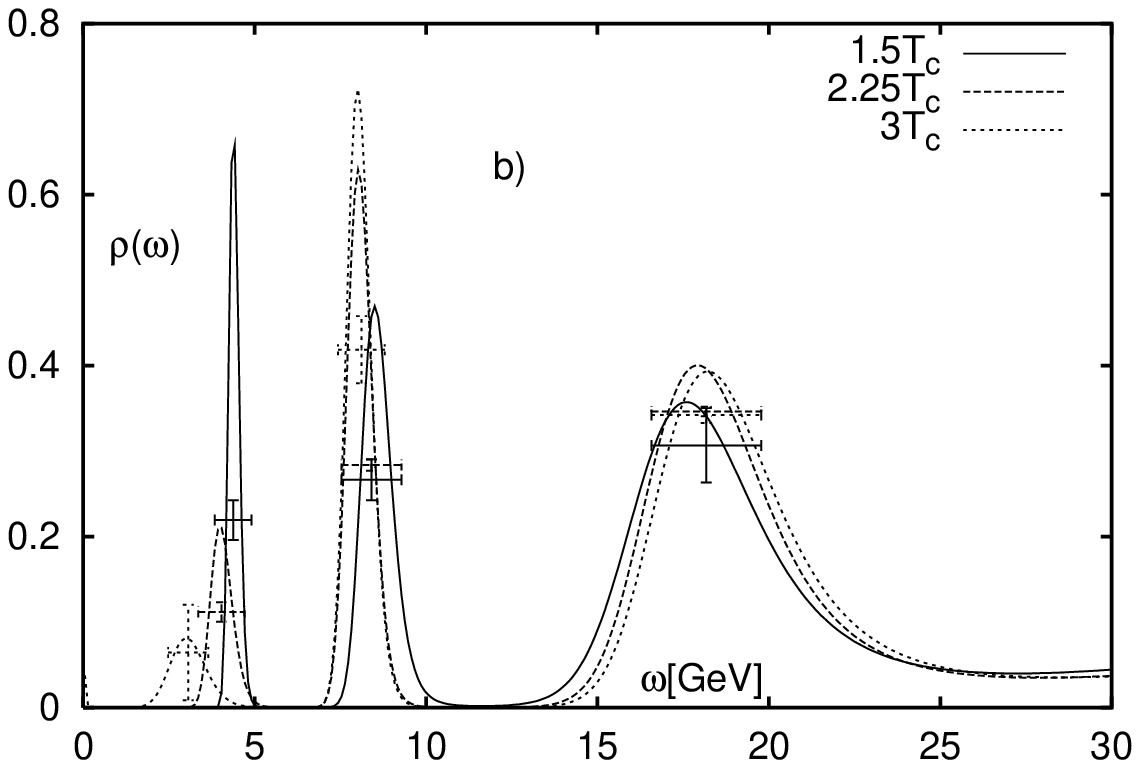}}
\caption{Spectral functions in the deconfined plasma, for a) \ec\ and
b) \jpsi\ channels.}
\label{fig.psvc719}
\end{figure} \end{center}

A more detailed view of the temperature modifications of
the mesons can be obtained by extracting the spectral function directly at
higher temperatures, by application of the MEM. At these
temperatures, the small extent of the temporal direction
makes a reliable extraction of the spectral function
difficult without proper a-priori information.
However, precise information about the
structure of \rhw\ at large $\om$, obtained from
\Fref{fig.psvc0.9} a), allows a reliable extraction of the
spectral function \cite{paper}. As part of the prior guess,
we provide the large $\om$ structure from 
\Fref{fig.psvc0.9} a), smoothly connected to $\sim \om^2$
behavior at low $\om$. The spectral functions extracted with
such a prior guess are shown in \Fref{fig.psvc719}.

\vspace*{-0.4cm}
\vspace*{-0.4cm}

\Fref{fig.psvc719} supports the trend seen in 
\Fref{fig.psvc0.9} b). In the \ec\ channel, a strong peak is
seen at 1.5 $\tc$, with an essentially unchanged position
and strength from that below $\tc$. 
In the \jpsi\ also a strong peak is seen upto 1.5
$\tc$; no significant reduction in mass or peak strength is
seen upto this temperature (see refernce \cite{paper} for a 
discussion of possible changes at this temperature). 
At 2.25 $\tc$, a significant peak is still seen in both 
the channels, but with a much reduced strength. Furthermore, we 
do not see a statistically significant peak at 3 $\tc$. 
(The error bars are standard
deviations over the integrated strength, integrated over
the marked region in the x direction; see reference \cite{jarrel}.)

\begin{center} \begin{figure} [htb]
\centerline{\includegraphics[width=6.2cm]{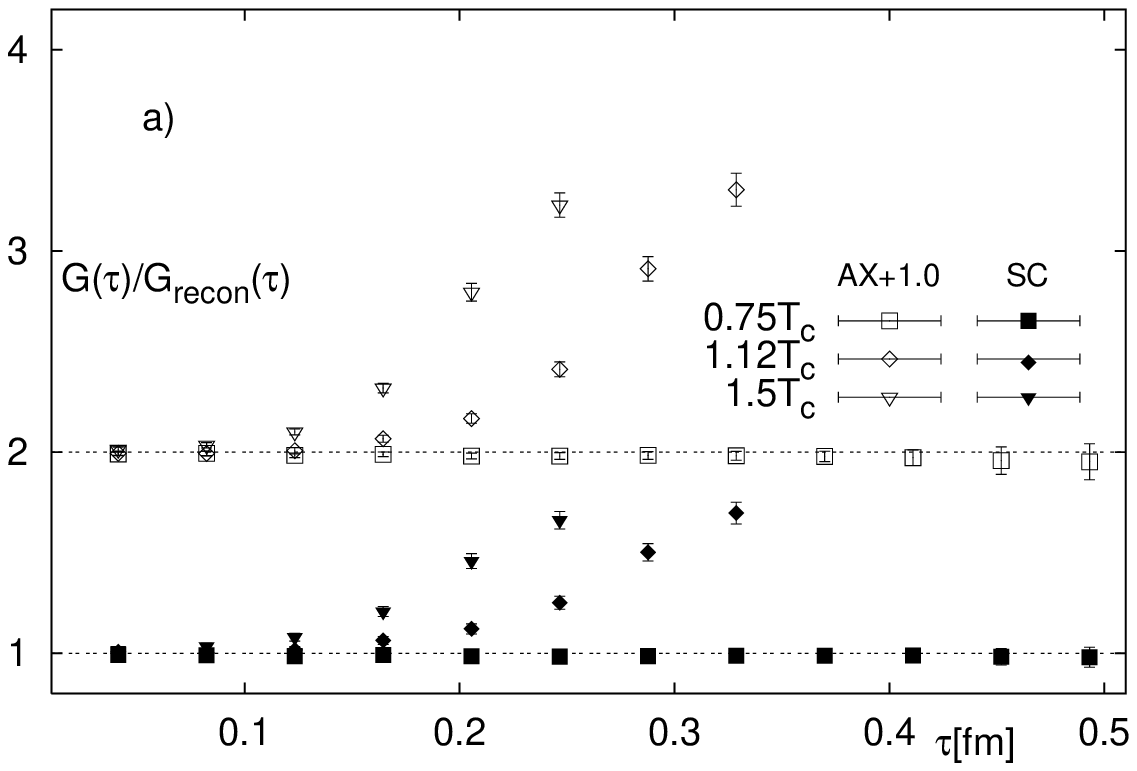}
\includegraphics[width=6.2cm]{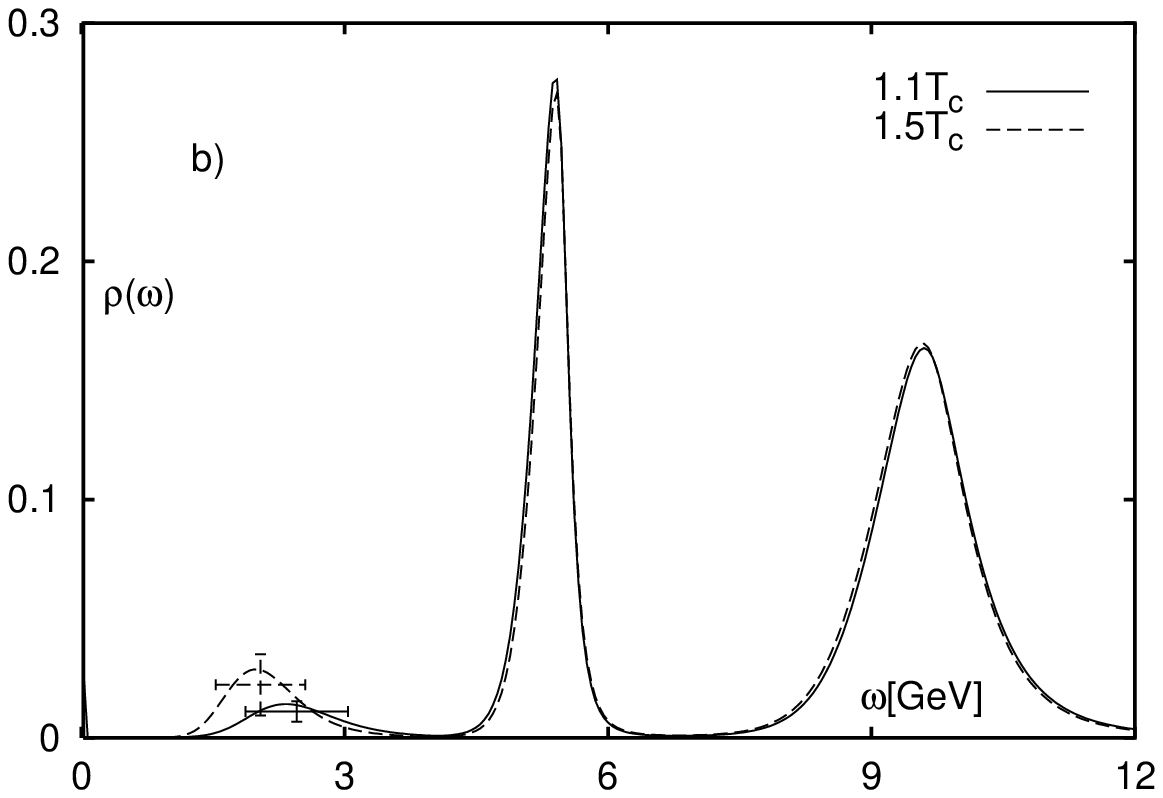}}
\caption{a) $\Gt / G_{\rm recon}(\tau, 0.75 \tc)$ for the
scalar and axial vector channels. b) The spectral
function for the scalar channel.}
\label{fig.scax664}
\end{figure} \end{center}

While the 1S states seem to be little affected by the
deconfinement transition, the situation seems to be quite
different for the 1P states \as\ and \ax\ . A similar
analysis indicates that these states are significantly
modified on crossing the transition point. In order to
study these states, we use the point
operators $J_H(x) = \bar{c} c$ and $\bar{c} \gamma_5 \gamma_\mu c$,
respectively. A reliable extraction of the spectral
function below $\tc$ is possible, 
with the ground state peak reproducing the mass and
strength obtained from a 
fit. At high $\om$, lattice artefacts similar to those in 
\Fref{fig.psvc0.9} a) dominate. In
\Fref{fig.scax664} a) we show the comparison of the
correlators for the axial vector and scalar channels at 1.1
and 1.5 $\tc$ with those reconstructed from the spectral function 
at 0.75 $\tc$. The figure indicates a very significant
modification of the properties of the states in these
channels, already at 1.1 $\tc$. \Fref{fig.scax664} b) shows
the spectral functions extracted above $\tc$ for the scalar 
channel. As above, we use the high energy structure below
$\tc$ as part of the prior guess. The figure indicates that
the \as\ state may have dissolved already at 1.1 $\tc$. 
A very similar figure is also obtained for the \ax\ state.

\vspace*{-0.3cm}
\vspace*{-0.4cm}

These results have direct experimental
relevance for the dilepton signal from an equilibriated plasma.
The differential dilepton rate is directly related to
\rhw\ in the vector channel:
\beq
{dW \over dw d^3p} |_{\vec{p}=0} = {5 \alpha^2 \over 27 \pi^2} 
{1 \over (e^{\om /T}-1)} \rho_V (\om, T).
\label{eq.dilepton} \eeq
Therefore our results for spectral function indicate that the
dilepton signal should have a significant \jpsi\ peak even
for plasma temperatures above 2 $\tc$. On crossing $\tc$,
the absence of the \chc\ states will cause a reduction of
the \jpsi\ peak due to the absence of the ``indirect''
\jpsi\ from \chc\ decay (this may happen even below $\tc$
for full QCD). No further significant reduction of the
peak strength will be seen at least upto temperatures of 1.5
$\tc$, while reaching even higher temperatures will result in
 a weakening of the peak (possibly due to collision broadening).
The results presented here, of course, only refer to a gluonic
plasma. However, in a deconfined quark-gluon plasma the
collision of the \jpsi\ with the hot thermal gluons is expected to be the
significant mechanism for dissolving the \jpsi\ 
\cite{kharzeev}; the above results thus should
hold at least qualitatively also for full QCD.

This research is supported by BMBF under grant no. 06BI102.

\end{document}